\documentclass{article}
\usepackage{amsmath,amssymb}

\newtheorem{thm}{Theorem}[section]

\newtheorem{lem}{Lemma}[section]

\newenvironment{proof}
{\noindent\textsc{Proof:}}{\hfill $\square$}

\newcommand{\V}{\mathbf{V}^1}
\newcommand{\Zp}{\mathbb{Z}_p}

\title{A polytime proof of correctness of the Rabin-Miller algorithm from
Fermat's Little Theorem}
\author{Grzegorz Herman\footnote{\tt hermang@mcmaster.ca} 
and Michael Soltys\footnote{\tt soltys@mcmaster.ca}}

\begin{document}
\maketitle
\begin{abstract}
Although a deterministic polytime algorithm for primality testing is
now known~(\cite{primes}), the Rabin-Miller randomized test of
primality continues being the most efficient and widely used
algorithm.  

We prove the correctness of the Rabin-Miller algorithm in the theory
$\V$ for polynomial time reasoning, from Fermat's little theorem.
This is interesting because the Rabin-Miller algorithm is a polytime
randomized algorithm, which runs in the class $\textbf{RP}$ (i.e., the
class of polytime Monte-Carlo algorithms), with a sampling space
exponential in the length of the binary encoding of the input number.
(The class {\bf RP} contains polytime {\bf P}.)  However,  we show how
to express the correctness in the language of $\V$, and we also show
that we can prove the formula expressing correctness with polytime
reasoning from Fermat's Little theorem, which is generally expected to
be independent of $\V$.

Our proof is also conceptually very basic in the sense that we use the
extended Euclid's algorithm, for computing greatest common divisors,
as the main workhorse of the proof.   For example, we make do without
proving the Chinese Reminder theorem, which is used in the standard
proofs.
\end{abstract}

\section{Introduction}

A deterministic polytime algorithm for primality testing is now
known~(\cite{primes}), although it does not follow that the
correctness of this algorithm can be shown with polytime concepts, and
it is not at all clear that there exists a polytime proof of
correctness.

In practice, the Rabin-Miller randomized algorithm for primality
testing is the most widely used algorithm.  It is fairly simple to
describe, and very efficient (in runs in time $O(n^4)$, where $n$ is
the size of the binary encoding of the input number).  The proof of
correctness is basic, in the sense that it does not use major results
of number theory; it is our task in this paper to provide a proof in
the polytime theory $\V$ (\cite{cook_phuong06}) from Fermat's Little
theorem (a ``pure'' $\V$ proof cannot be expected as Rabin-Miller is
an {\bf RP} algorithm).  Our result distills the hard (from a proof
complexity point of view) theorem behind the correctness of the
algorithm.

The proof complexity of randomized algorithm has been studied in depth
in~\cite{jerabek}, and indeed it is shown
there~(\cite[Example~3.2.10]{jerabek}) that there is an {\bf RP}
predicate $P(x)$, which is $1/2$-definable in Buss' polytime theory
$\mathbf{S}_2^1$, such that $\mathbf{S}_2^1$ proves ``$P(x)\text{ iff
}\text{Fermat's Little Theorem}$''.  In our case, we use the basic
machinery of $\V$ and the following assertion of correctness: for
every non-witness of compositness there is a unique witness of
compositness (see Figure~\ref{fig:1}).  This shows that at least half
the elements of the sample space are witnesses, and proves the
correctness of the algorithm.

Further, \cite{jerabek} claims that $\mathbf{S}_2^1$ is able to prove
that every number is uniquely representable as a product of prime
powers--- and the proof of the correctness of the Rabin-Miller
algorithm relies on this.  If we could prove the same fact in $\V$, we
would have a polytime algorithm for factoring.  This is the main
difference when using our technique; we never argue about
factorization of numbers.  The $1/2$-definability (or, in general,
$s/t$-definability) given in~\cite{jerabek} is a slightly more general
approach to comparing set sizes. To state that $|A|$ is at least
$(s/t)|B|$, it states the existence of a surjective mapping from
$t\cdot A$ to $s\cdot B$.  In this paper, we force our mapping to be
multiplication modulo $P$, whereas~\cite{jerabek} makes it any
polysize circuit.

No extra assumptions are necessary to prove the correctness of the
algorithm on composites.  However, to show that there are no false
negatives, i.e., to show that the algorithm always answers correctly
on inputs that {\em are} prime numbers, we use Fermat's little
theorem.

While there is no independence result showing that $\V\nvdash$
``Fermat's little theorem'', it is believed that it is not provable in
$\V$.  The reason for this belief (following~\cite{cook_phuong06}) is
that the existential content of Fermat's little theorem can be
captured by its contrapositive form:
\begin{equation}\label{eq:1}
\underbrace{(1\le a<n)\wedge(a^{n-1}\neq 1\pmod
n)}_{\text{hypothesis}} \supset\exists d(1<d<n\wedge d|n)
\end{equation}
and if we could prove Fermat's theorem in $\V$, we could obviously
prove the above formula as well (note that $a^{n-1}\pmod n$ can be
computed in polytime by repeated squaring).  

If~(\ref{eq:1}) were provable in $V$, then by a witnessing theorem it
would follow that a polytime function $f(a,n)$ exists whose value
$d=f(a,n)$ provides a proper divisor of $n$ whenever $a,n$ satisfy the
hypothesis of~(\ref{eq:1}).  With the exception of the so-called
Carmichael numbers, which can be factored in polynomial time, every
composite $n$ satisfies the hypothesis for at least half of the values
of $a$, $1\le a<n$.  Hence, $f(a,n)$ would provide a probabilistic
polytime algorithm for integer factoring.  Such an algorithm is
thought unlikely to exist, and would provide a method for breaking the
RSA public-key encryption scheme.

In short, it is interesting to see how strong a theory one needs in
order to prove the correctness of the Rabin-Miller algorithm.  Since
we do not know if it is possible to derandomize probabilistic polytime
computations, we cannot hope to have a purely polytime proof in this
case.  It is still worthwhile to isolate the assumptions on which the
theory ``falls short'' of the task, i.e., what is the principle
underlying the Rabin-Miller algorithm which is responsible for the
apparent inability of a polytime theory to prove its correctness?  We
answer that it is the Fermat's little theorem, and show that $\V$
proves the equivalence of the correctness of Rabin-Miller algorithm
(properly stated) and Fermat's Little theorem.

This paper is organized as follows.  In section~\ref{sec:v} we
describe very briefly the theory $\V$ for polytime reasoning.  For a
full background on $\V$ see the book~\cite{cook_phuong06}.  In
section~\ref{sec:nt} we give some number theoretic preliminaries, we
recall extended Euclid's algorithm, and say that it can be shown
correct in $\V$.  We also recall Euler's theorem, and Fermat's Little
theorem.  In section~\ref{sec:pseudo} we show how we can build an
algorithm for pseudoprimality (a number is pseudoprime if it is prime
or a Carmichael number) from Fermat's Little theorem.  This introduces
the Rabin-Miller test of primality, which extends the pseudoprimes by
coping with the Carmichael numbers.  The presentation of the
Rabin-Miller algorithm, and its $\V$ proof of correctness from
Fermat's Little theorem, are presented in section~\ref{sec:r-m}.

Finally, note that the original work on the Rabin-Miller algorithm has
been published in~\cite{miller,rabin}, but we use the presentation of
the algorithm as given in~\cite{sipser}.

\section{The theory $\V$}\label{sec:v}

In this section we introduce briefly the theory $\V$ for polytime
reasoning; see~\cite{cook_phuong06} for a full and detailed treatment.

$\V$ is a two sorted theory, where the two sorts are indices and
strings.  The strings are formally sets of numbers, where the
correspondence with strings is given by $i\in X$ iff the $i$-th bit is
1.  We think of the strings as numbers encoded in binary.  The indices
are unary numbers used to index the strings, and their role is
auxiliary; the main objects of interest are strings, which will encode
numbers.  The vocabulary of our theory is
$\mathcal{L}_A^2=[0,1,+,\cdot,||;=_1,=_2,\le,\in]$.

Here the symbols $0,1,+,\cdot,=_1$ and $\le$ are from the usual
vocabulary of Peano Arithmetic, and they are function and predicate
symbols over the first sort (indices).  The function $|X|$ (the
``length of $X$'') is a number-valued function and it intended to
denote the length of the string $X$.  The binary predicate $\in$ takes
a number and a string as arguments, and is intended to be true if the
position in the string given by this number is 1.  (Note that
technically, the strings are sets of numbers; hence the set theoretic
notation.)  Finally, $=_2$ is the equality predicate for the
second-sort objects.  We will write $=$ for both $=_1$ and $=_2$, and
which one it is will be clear from the context.
Sometimes we shall use the abbreviation $X(t)$ for $t\in X$.

We denote by $\Sigma_0^B$ the set of formulas over the language
$\mathcal{L}_A^2$ whose only quantifiers are bounded number
quantifiers, and we denote by $\Sigma_1^B$ the set of formulas of the
form 
$$
(\exists X_1\le t_1)\cdots(\exists X_n\le t_n)\alpha
$$ 
where $\alpha$ is a $\Sigma_0^B$ formula.  Here the expression
$(\exists X\le t)$ denotes $(\exists X)[|X|\le t]$.

\begin{figure}[h]
\begin{tabular}{ll}
B1.  & $x+1\neq 0$ \\
B2.  & $x+1=y+1\supset x=y$ \\
B3.  & $x+0=x$ \\
B4.  & $x+(y+1)=(x+y)+1$ \\
B5.  & $x\cdot 0=0$ \\
B6.  & $x\cdot(y+1)=(x\cdot y)+x$ \\
B7.  & $(x\le y\wedge y\le x)\supset x=y$ \\
B8.  & $x\le x+y$ \\
B9.  & $0\le x$ \\
B10. & $x\le y\vee y\le x$ \\
B11. & $x\le y\leftrightarrow x<y+1$ \\
B12. & $x\neq 0\supset\exists y\le x(y+1=x)$ \\
L1.  & $y\in X\supset y<|X|$ \\
L2.  & $y+1=|X|\supset y\in X$ \\
SE.  & $[|X|=|Y|\wedge\forall i<|X|(i\in X\leftrightarrow i\in Y)]
       \supset X=Y$
\end{tabular}
\caption{The 2-BASIC axioms.}\label{fig:2}
\end{figure}

For a set of formulas $\Phi$, the {\em Comprehension Axiom Scheme},
$\Phi$-COMP, is the set of formulas
$$
(\exists X\le y)(\forall z<y)(X(z)\leftrightarrow\phi(z))
$$
where $\phi(z)$ is any formula in $\Phi$, and $X$ does not occur free in
$\phi(z)$.

The theory $\mathbf{V}^i$, for $i=0,1$ is the theory with the axioms
2-BASIC (in figure~\ref{fig:2}) and the $\Sigma_i^B$-COMP axiom
scheme.

Proving the correctness of the Rabin-Miller algorithm we are going to rely
heavily on the following theorem, proved in~\cite{cook_phuong06}:

\begin{thm}[$\V$ captures polytime reasoning]
A function $f: \{0,1\}^* \to \{0,1\}^*$, i.e., $f$ is a function from
strings to strings, is polytime computable iff
there exists a formula $\phi\in\Sigma_1^B$ such that:
\begin{align*}
  & \phi(X,Y) \iff f(X)=Y \\
  & \V \vdash \forall X \exists Y \phi(X,Y)
\end{align*}
\end{thm}

See~\cite{cook_phuong06} for a proof of this theorem.

The theory $\V$ allows us to prove induction and minimization axioms
from the axioms we already have.  As we make use of those in our proof
of the correctness of the Rabin-Miller algorithm, we state them here
explicitly.

The {\em Number Induction Axiom} states that if $\Phi$ is a set of
two-sorted formulas, then $\Phi$-IND axioms are the formulas
$$
[\phi(0)\wedge\forall x,\phi(x)\supset\phi(x+1)]\supset\forall
z\phi(z)
$$
where $\phi$ is a formula in $\Phi$.

The {\em Number Minimization Axiom} states that if $\Phi$ is a set of
two-sorted formulas, then $\Phi$-MIN axioms are the formulas
$$
\exists z\phi(z)\supset\exists y[\phi(y)\wedge\neg\exists
x(x<y\wedge\phi(x))]
$$ 
where $\phi$ is a formula in $\Phi$.

We are of course interested in the cases where $\Phi$ is either
$\Sigma_0^B$ or $\Sigma_1^B$.

\begin{thm}
For $i=0$ or $i=1$, $\mathbf{V}^i$ proves both $\Sigma_i^b$-IND and
$\Sigma_i^b$-MIN.
\end{thm}

See~\cite{cook_phuong06} for a proof of this theorem.  Note that this
theorem allows us to do induction of $\Sigma_1^B$ formulas, and
minimization over $\Sigma_1^B$ formulas, when arguing about the
correctness of the Rabin-Miller theorem, without taking us outside the
polytime theory $\V$.

\section{Number theoretic background}\label{sec:nt}

In this section we give the basic number theoretic notions that will
be used in our paper, as well as recall Euler's theorem and its
corollary, Fermat's Little theorem.

We do not need Euler's theorem in our proof of correctness, but we
include it since it provides the most general proof of Fermat's Little
theorem which is the principle from which, as we show, the correctness
of the Rabin-Miller algorithm follows.  We recall that Euler's theorem
itself follows directly from Lagrange's theorem (of course, it also
follows directly from the Prime Factorization theorem).

We also present Euclid's algorithm for computing the greatest common
divisor of two numbers.  The correctness of the extended Euclid's
algorithm (provable in $\V$) is the main workhorse of our proof.   

Two numbers $x,y$ are {\em equivalent modulo} a third number $p$ (we
write $x=y\pmod p$) if they differ by a multiple of $p$. Every number
is equivalent modulo $p$ to some number in $\Zp=\{0,1,\ldots,(p-1)\}$.

For convenience we let $\Zp^+=\{1,\ldots,(p-1)\}$. We let $\Zp^*$ be
the subset of $\Zp^+$ of elements $a$ such that $\gcd(a,p)=1$. Note
that $(\Zp,+)$ is a group (under addition) and $(\Zp^*,\cdot)$ is a
group (under multiplication). The latter fact means that $\Zp^*$ can
be alternatively defined as
$$
\{a\in\Zp^+|\text{$a$ has a (multiplicative) inverse in $\Zp^+$}\}
$$
and it follows from the next lemma.

\begin{lem}[Euclid's Lemma]
For any two numbers $a$ and $b$ there exist numbers $x$ and $y$ such
that $ax+by = \gcd(a,b)$.  Furthermore, the correctness of the
extended Euclid's algorithm (where ``correctness'' simply states that
on input $a,b$ the output $x,y$ satisfies the condition
$ax+by=\gcd(a,b)$) is provable in $\V$.
\end{lem}

\begin{proof}
The lemma can be proved by analyzing the extended Euclid's algorithm:

\begin{figure}[h]
\begin{tabbing}
On \= input $(a,b)$: \\
   \> 1. if \= $a<b$ then \\
   \> 2.    \> let $(y,x,d) := \text{euclid}(b,a)$ \\
   \> 3.    \> return $(x,y,d)$ \\
   \> 4. if \= $b=0$ then \\
   \> 5.    \> return $(1,0,a)$ \\
   \> 6. let $(z,x,d) := \text{euclid}(b,a \mod b)$ \\
   \> 7. return $(x,z-(a \div b)x,d)$
\end{tabbing}
\caption{Extended Euclid's algorithm}\label{fig:4}
\end{figure}

The correctness of the algorithm is easily shown by induction, with the
inductive step (for lines 6-7) proved as follows:
\begin{align*}
  ax + b(z - (a \div b)x) &= ax + bz - b(a \div b)x \\
                          &= bz + (a-b(a \div b))x \\
                          &= bz + (a \mod b)x \\
                          &= d
\end{align*}
This is clearly a proof that can be carried out in polynomial time,
i.e., in $\V$.
\end{proof}

The easiest way to prove Euler's theorem is from Lagrange's theorem.
The proof of Lagrange's theorem is basic, and it is included in all
standard algebra textbooks.  Still, it is a proof that we do not know
how to carry out in $\V$.

\begin{thm}[Lagrange's Theorem]
If $H$ is a subgroup of $G$, then the order of $H$ divides the order
of $G$, i.e., $H\le G$ $\Rightarrow$ $|H|||G|$.  In particular, the
order of any element divides the order of the group.
\end{thm}

The function $\phi(n)$ is called the {\em Euler totient function}, and
it is the number of elements less than $n$ that are co-prime to $n$,
i.e., $\phi(n)=|\mathbb{Z}_n^*|$.   If we are able to factor, we are
also able to compute $\phi(n)$: suppose $n=p_1^{k_1}p_2^{k_2}\cdots
p_l^{k_l}$, then $\phi(n)=\prod_{i=1}^lp_i^{k_i-1}(p_i-1)$.

\begin{thm}[Euler's Theorem]
For every $n$ and every $a\in\mathbb{Z}_n^*$, $a^{\phi(n)}=1\pmod n$.
\end{thm}

\begin{proof}
This is a consequence of Lagrange's Theorem (which says that the order
of any subgroup, and hence the order of any element, divides the order
of the group).
\end{proof}

\begin{thm}[Fermat's Little Theorem]
For every prime $p$ and every $a\in\Zp^+$, we have $a^{(p-1)}=1\pmod p$.
\end{thm}

\begin{proof}
A consequence of Euler's Theorem.  Note that when $p$ is a prime,
$\Zp^+=\Zp^*$, and $\phi(p)=(p-1)$.
\end{proof}

Currently we do not have a polytime proof of Fermat's Little theorem,
and for the reasons outlined in the introduction we do not expect to
be able to prove it in a theory like $\V$, since a standard witnessing
argument would then imply that we can have a randomized polytime
algorithm for factoring, which is something that is generally not
believed to be possible.

As an aside, note that a stronger induction than the one in $\V$,
i.e., an induction that can be carried out on ``values'' of strings,
rather than on ``notation'', which means an induction of the kind as
in the theory $\mathbf{T}_2^1$ (see~\cite[\S 5.2]{krajicek}), can
prove Fermat's Little theorem.  Here is the outline of the proof: we
show that for $\gcd(a,p)=1$, $a^p=a\pmod p$, by induction on $a$.  It
is enough to prove this, since if $\gcd(a,p)=1$, then $a$ has an
inverse in $\mathbb{Z}_p^+$, and so Fermat's Little theorem follows.
The basis case is trivial: $1^p=1\pmod p$.  Now
$(a+1)^p=a^p+1+\sum_{j=1}^{p-1}\binom{p}{j}a^{p-j}$ (where we need
$\Sigma_1^B$ formulas to express the binomial expansion).  Note that 
$\sum_{j=1}^{p-1}\binom{p}{j}a^{p-j}=0\pmod p$, and so the result
follows.

\section{Pseudoprimes}\label{sec:pseudo}

Fermat's little theorem provides a ``test'' for primality, called the Fermat
test.  When we say that $p$ passes the Fermat test at $a$, we mean that
$a^{(p-1)}=1\pmod p$.  Thus, all primes pass the Fermat test for all
$a\in\Zp^+$.

Unfortunately, there are also composite numbers $n$ that pass the Fermat
tests at every $a\in\mathbb{Z}_n^*$; these are the so called {\em Carmichael
numbers}
%
%\footnote{(From {\tt Wolfram MathWorld}) R.\ D.\ Carmichael first
%noted the existence of such numbers in 1910, computed 15 examples,
%and conjectured that though they are infrequent there were infinitely
%many.  In 1956, Erdos sketched a technique for constructing large
%Carmichael numbers~(\cite{lovednumbers}), and a proof was given
%by~\cite{carmichael} in 1994.} 
%
(e.g., 561, 1105, 1729).

\begin{lem}
If $p$ is a composite non-Carmichael number, then it passes Fermat's
test for at most half of the elements of $\Zp^*$.
\end{lem}

\begin{proof}
(This is exercise~10.16 in~\cite{sipser})
Call $a$ a {\em witness} if it fails the Fermat test for $p$, that is, if
$a^{(p-1)}\neq 1\pmod p$.

Consider $S\subseteq \Zp^*$ consisting of those elements $a\in\Zp^*$ for which
$a^{p-1}=1\pmod p$.  It is easy to check that $S$ is in fact a {\em subgroup}
of $\Zp^*$. Therefore, using the Lagrange Theorem, $|S|$ must divide $|\Zp^*|$.
Suppose now that there exists an element $a\in\Zp^*$ for which
$a^{p-1}\neq 1\pmod p$.  Then, $S$ is not ``everything'' (i.e., not $\Zp^*$),
so the next best thing it can be is ``half'' (of $\Zp^*$).
\end{proof}

A number is {\em pseudoprime} if it is either prime or Carmichael. The last
lemma suggests an algorithm for pseudoprimes: on input $p$, check
$a^{(p-1)}=1\pmod p$ for some random $a\in\Zp^+$.  If the test fails (i.e.,
$a^{(p-1)}\neq 1$), then $p$ is composite for sure. If $p$ passes the test,
then it is probably pseudoprime. From the above lemma we know that the
probability of error in this case is $\le\frac{1}{2}$.  Note that if
$\gcd(a,p)\neq 1$, then $a^{(p-1)}\neq 1\pmod p$.  Thus, on Carmichael
numbers, the algorithm for pseudoprimness might answer sometimes
``composite'', and sometimes ``pseudoprime''.

\section{Rabin-Miller Algorithm}\label{sec:r-m}

The Rabin-Miller algorithm (Figure~\ref{fig:3}) ``copes'' with the
Carmichael numbers, in effect turning the algorithm for
pseudoprimality given in the previous section into an algorithm for
primality.

\begin{figure}[h]
\begin{tabbing}
On \= input $(p,a)$: \\
   \> 1. \= If $p$ is even, accept if $p=2$; otherwise, reject. \\
   \> 2. \> Compute $a^{(p-1)}\pmod p$ and reject if $\neq 1$.\\
   \> 3. \> Let $(p-1)=s2^h$ where $s$ is odd. \\
   \> 4. \> Compute the sequence \\
   \>    \> $a^{s\cdot 2^0},a^{s\cdot 2^1},a^{s\cdot 2^2},
            \ldots,a^{s\cdot 2^h}\pmod p$. \\
   \> 5. \> If some element of this sequence is not $1$, \\
   \>    \> find the last element that is not $1$, \\ 
   \>    \> and reject if that element is not $-1$. \\
   \> 6. \> Accept.
\end{tabbing}
\caption{The Rabin-Miller algorithm.}\label{fig:3}
\end{figure}

Note that if we got to line 4. in the algorithm, it means that
$a^{s\cdot 2^h}=1\pmod p$.  We say that $a$ is a {\em witness} (of
compositness) of type 1 or type 2 if the algorithm rejects at step 2
or step 5, respectively.

The algorithm is polytime (we can compute the sequence in step 4 via
iterated squaring). If we randomly select the $a$ from $\Zp^+$, it
will become a {\bf RP} algorithm.

Before proving that the algorithm is correct, we have to state this
fact in the language of our theory. We would like to say that ``there
are few false positives''. The meaning of ``few'' can be chosen to be
``at most one half'' (if we need a better bound, we can achieve them
using the idea of {\em amplification}, meaning that we repeat the
algorithm $k$ many times, on independently selected $a$'s, and achieve
an error of $\frac{1}{2^k}$; which for $k$ equal to, say, 100, is
negligible). 

But how do we speak about probability? The obvious way would be to
express our event space and capture the size of the subset of ``bad''
events (i.e., the non-witnesses). But this is not possible in $\V$,
because the event space is exponential in length of the input $P$,
and $\V$ only allows us to talk about polynomial-length strings (and
giving it more power in this domain would allow us to capture more
than polytime reasoning and thus defeat the purpose of this analysis).

How then can we compare the cardinalities of two sets without
mentioning them explicitly? The set of non-witnesses is at most half
of the size of the set of all candidates if and only if there exists
an injective mapping from non-witnesses to witnesses. Again, stating
an existence of such a mapping in general is not possible in $\V$, so
we strengthen our goal to prove the existence of a particular type of
mapping---see figure~\ref{fig:1}.
\begin{figure}[h]
\begin{tabbing}
$1<D<P$ \= $\land \,\,\, D | P$ $\implies$ \\
        \> $\exists$ \= $T,T'\le|P|$ such that \\
        \>           \> $T*T' = 1\pmod P$ and \\
        \>           \> $\forall$ \= $A\le|P|$ \\
        \>           \>           \> ``$A$ is a non-witness'' \\
        \>           \>           \> $\implies$ ``$(A*T\pmod P)$ is a witness''
\end{tabbing}
\caption{Correctness assertion.}\label{fig:1}
\end{figure}
Because we require $T$ to have an inverse $T'$ modulo $P$, we know that the
function mapping $A$ to $A*T$ is injective. Note that the statement we want to
prove is not a $\Sigma_1^B$ formula. But this is not a problem, as $\V$ only
restricts the comprehension axiom scheme (and thus the induction) to
$\Sigma_1^B$ formulas.

We will start by showing that a composite $P$ is either a power of a
smaller number $Q$, or a product of two relatively prime numbers $Q$
and $R$. Because we do not know how to talk about prime factorization
of $P$ in $\V$, we will use the following recursive algorithm:

\begin{figure}[h]
\begin{tabbing}
On \= input $(Q,E,R)$: \\
   \> 1. while \= $\gcd(Q,R) = G > 1$ \\
   \> 2.       \> if $G = Q$, let $(Q,E,R) := (Q,E+1,R/Q)$ \\
   \> 3.       \> otherwise, let $(Q,E,R) := (Q/G,E,G^ER)$ \\
   \> 4. return $(Q,E,R)$
\end{tabbing}
\caption{Factoring.}\label{fig:5}
\end{figure}

It is not difficult to see that the while loop preserves the following
invariants:
\begin{itemize}
\item $P=Q^ER$
\item $Q>1$
\item $R=1 \implies E>1$
\end{itemize}
Therefore the result gives us either $P=Q^E$ with $E>1$ (when $R=1$),
or $P=QR$ with $Q,R>1$ and $\gcd(Q,R)=1$. Moreover, every iteration
either increases $E$ by $1$ or decreases $Q$ by at least half, so the
algorithm runs in polynomial time. Therefore, given that $P$ is
composite, and we have a factor $D$ of $P$, i.e., $1<D<P, D|P$, we can
initialize the algorithm with $(D,1,P/D)$ and thus prove (in $\V$)
that one of two desired cases holds indeed.

In the case when $P=Q^E, E>1$, we simply set
\begin{align*}
  T  &:= 1+Q^{(E-1)}, \\
  T' &:= T^{(P-1)} \pmod P.
\end{align*}
Then we can show (by induction on the length of $J$) that
$$
  T^J = 1 + JQ^{(E-1)} \pmod P,
$$
and conclude that
$$
TT' = T^P = 1 \pmod P.
$$
Moreover, whenever $A$ is a non-witness, we know that 
$$
A^{(P-1)}=1 \pmod P,
$$ 
and thus 
$$
(AT)^{(P-1)} = T^{(P-1)} = T' \neq 1 \pmod P,
$$ 
so $AT$ is a (type 1) witness, as required.

In the other case more work needs to be done. First we represent
$(P-1) = S2^h$, with odd $S$, as in the algorithm. Then we let
$$
  \alpha(i) := (\exists Z\le|P|)[Z^{S2^i}=-1 \pmod P].
$$
From the fact that $S$ is odd we know that $\alpha(0)$ (take $Z=P-1$).
Now $\alpha(h)$ is either true or false. If it is true, then we let
both $T$ and $T'$ to be the $Z$ witnessing that fact. Thus we have:
$$
  TT' = Z^2 = (-1)^2 = 1 \pmod P,
$$
and, as before, whenever $A$ is a non-witness, $AT$ is a (type 1) witness.

When $\alpha(h)$ if false then by minimality principle (equivalent to
induction, and allowed because $\alpha$ is a $\Sigma_1^B$ formula) we
can get the smallest $i$ for which $\alpha(i+1)$ is false. Let $Z$ be
the witness of $\alpha(i)$ being true.  Remember that we have a
factoring $P=QR$, with $\gcd(Q,R)=1$. According to Euclid's lemma we
can compute $X$ and $Y$ such that
$$
  XQ + YR = \gcd(Q,R) = 1.
$$
Now we let $T := XQ + YZR \pmod P$, $T' := T^{S2^{i+1}-1}$ and notice that
\begin{align*}
  T        &= XQ + YZR \\
           &= XQ + YZR + X(Z-1)Q \\
           &= Z (XQ + YR) = Z            &\pmod Q \\
  T        &= XQ + YZR \\
           &= XQ + YZR - Y(Z-1)R \\
           &= XQ + YR = 1                &\pmod R \\
  T^{S2^i} &= Z^{S2^i} = -1              &\pmod Q \\
  T^{S2^i} &= 1^{S2^i} = 1               &\pmod R \\
  TT'      &= T^{S2^{i+1}} = (-1)^2 = 1  &\pmod Q \\
  TT'      &= T^{S2^{i+1}} = 1^2 = 1     &\pmod R \\
  TT'      &= 1                          &\pmod P
\end{align*}

Suppose that $P|(T^{S2^i}+1)$. Then $R|(T^{S2^i}+1)$. But as
$R|(T^{S2^i}-1)$, we would have that 
$$
R|((T^{S2^i}+1)-(T^{S2^i}-1))=2
$$ 
and thus $2=R|P$ which is
not possible, as the algorithm deals with even $P$'s in step 1. 

Analogously, we cannot have $P|(T^{S2^i}-1)$. Therefore we know that
$T^{S2^i}\neq\pm 1 \pmod P$. Now, if we consider any non-witness $A$,
we will have 
$$
A^{S2^i}=\pm 1 \pmod P \text{ and } A^{S2^{i+1}}=1 \pmod P
$$ 
owing the way $i$ was chosen. But then $(AT)^{S2^i}\neq\pm 1 \pmod P$
and $(AT)^{S2^{i+1}}=1 \pmod P$, so again $AT$ is a (type 2) witness.

Having considered all the cases, we have proved (in $\V$) that the
probability of accepting a composite number is at most $\frac{1}{2}$.
To arrive at the correctness of the Rabin-Miller test we need to prove
one last lemma:

\begin{lem}
Suppose that $P$ is a prime number. Then the Rabin-Miller algorithm
accepts $(P,A)$ for every $A \in \Zp^+$ (that is, there are no false
negatives).
\end{lem}

\begin{proof}
Assume that $P$ is prime, but the algorithm rejects $(P,A)$. If $A$
was a type~1 witness, $A^{(P-1)}\neq1\pmod P$ then Fermat's little
theorem would imply that $P$ is composite. If $A$ was a type 2
witness, some $B$ exists in $\Zp^+$, where $B\neq\pm 1\pmod P$ and
$B^2=1\pmod P$. Therefore, $(B^2-1)=0\pmod P$, and so $P$ has to
divide $(B-1)(B+1)$.  But because $B\neq\pm1\pmod P$, both $(B-1)$ and
$(B+1)$ are strictly between $0$ and $P$.  As we assumed $P$ to be a
prime, we have $\gcd(P,B-1)=\gcd(P,B+1)=1$, and (using Euclid's
lemma), $\gcd(P,(B-1)(B+1))=1$, a contradiction.
\end{proof}

The only part of this lemma (and thus of the whole proof of
correctness) not shown in $\V$ is the Fermat's little theorem. It is
also obvious that it is implied by the correctness of the Rabin-Miller
algorithm. Therefore we can formulate the main result of this work:

\begin{thm}
$\V$ proves the equivalence of Fermat's little theorem to the
correctness of the Rabin-Miller randomized algorithm for primality.
\end{thm}

\section{Conclusion}

We gave a direct and conceptually simple proof of the equivalence, in
$\V$, of the correctness of the Rabin-Miller theorem (properly
stated), and Fermat's Little Theorem.  The proof relies on rudimentary
number theory, and more concretely, on a proof of correctness in $\V$
of the extended Euclid's algorithm for computing the greatest common
divisor.  

It is a very interesting open problem, although probably very
difficult, to show an independence of Fermat's Little theorem from
$\V$, and hence the independence of the correctness of the
Rabin-Miller algorithm from $\V$.

%\bibliographystyle{abbrv}
%\bibliography{herman_soltys_rm}

\end{document}